\documentclass[12pt]{article}
\usepackage{cmp2e}
\usepackage{graphicx}
\usepackage{amssymb}
\usepackage{epstopdf}

\title{Interaction Between Two Rows of Localized Adsorption Sites in a 2D One-Component Plasma}
\author{Christian D. Santangelo\refaddr{Penn}, Lesser Blum\refaddr{PuertoRico}}
\addresses{
\addr{Penn} Department of Physics, University of Pennsylvania, Philadelphia, 19104
\addr{PuertoRico} Department of Physics, Faculty of Natural Sciences, University of Puerto Rico, Rio Piedras, Puerto Rico 00931
}

\begin{document}

\maketitle
%\password: ssbmd

%\title{Interaction Between Two Rows of Localized Adsorption Sites in a 2D One-Component Plasma}

%\author{Christian D. Santangelo}
%\email{santancd@physics.upenn.edu}
%\affiliation{Department of Physics and Astronomy, University of Pennsylvania, Philadelphia, 19104}
%\author{Lesser Blum}
%\email{lesblum@rrpac.upr.clu.edu}
%\affiliation{Department of Physics, Faculty of Natural Sciences, University of Puerto Rico, Los Piedras, Puerto Rico 00931}
%\date{\today}

\begin{abstract}
We compute the free energy for two rows of localized adsorption sites embedded in a two dimensional one-component plasma with neutralizing background density $\rho$.  The interaction energy between the adsorption sites is repulsive.  We also compute the average occupation number of the adsorption sites and compare it to the result for a single row of sites. The exact result indicates that the discretization does not induce charge asymmetry and no attractive forces occur.
\end{abstract}

The subtleties of electrostatics in condensed matter theory represent a formidable and never ending challenge. One  topic of much recent activity, has been  the attraction between two macromolecules of the same charge~\cite{electrostaticsreviews}. One mechanism that has been proposed invokes charge asymmetry related to the formation of lattices or Wigner crystals \cite{arup,bloomfield1}. One problem with this picture is that it will create a dipole that is inconsistent with the perfect screening sum rules (Blum \textit{et al.}\cite{perfscr}). While the formation of Wigner crystals under special conditions is an experimental fact, the question of the large asymmetry in the charge distribution needs clarification. As has happened in the past the exact solution of the two dimensional Jancovici model ~\cite{jancovici} can provide an unambiguous answer to the puzzle. The interaction between two equally charged lines ( which are charged surfaces in 2 dimensions!) has been discussed using the exact solution, and is always repulsive \cite{jancovici2}. 
Here, we study a discretized version of this problem, namely two lines of discrete adsorption sites, where the adsorption potential is given by  the Baxter~\cite{baxter} sticky potential. To do this we extend the localized adsorption model of a single line ~\cite{localizedadsorption} to the case of two lines of discrete adsorption sites. This extension is non-trivial, and as in a  similar case discussed in the past, has a simple solution for what we would call a 'commensurate'  lattice \cite{cornu}, namely the spacing of the adsorption sites is such that the background charge of the enclosed area corresponds to an entire number of discrete charges.

\section{Formalism}

\subsection{Modeling the Adsorption}
Following Rosinberg \textit{et al.}~\cite{localizedadsorption}, the adsorption potential for a sticky site located at the origin, given by $u_a(\textbf{r})$, is modeled as
\begin{equation}
\exp \left[ - \beta u_a(\textbf{r}) \right] = 1 + \lambda \delta(\textbf{r}),
\end{equation}
where $\lambda$ is a positive constant that measures the strength of the adsorption potential~\cite{baxter}.

The partition function for a system of adsorption sites with locations given by the vectors $\textbf{R}_m$ is given by
\begin{equation}
Z_N = \frac{1}{N!} \int e^{-\beta V_0(\textbf{r}_1, \cdots, \textbf{r}_N)} \prod_{i=1}^N \left\{ \left[1+\lambda \sum_{m=1}^M \delta(\textbf{r}_i - \textbf{R}_m) \right] d^2r_i \right\},
\end{equation}
where $V_0$ is the potential energy of the one-component plasma in the absence of adsorption sites.  Expanding in powers of $\lambda$, it has been shown~\cite{localizedadsorption} that the partition function can be written in terms of the $n$-point correlation functions as
\begin{equation}
Z_N = Z_N^0 \sum_s \sum_{m1, m2, \cdots m_s =1}^M \frac{\lambda^s}{s!} \rho(\textbf{R}_{m_1}, \cdots \textbf{R}_{m_s}),
\end{equation}
where $Z_N^0$ is the partition function of the unperturbed system.

The difference in free energy from the unperturbed system is the logarithm of $Z_N/Z_N^0$, and is given by
\begin{equation}
\label{eq:freeenergysum}
\Delta F = - k_B T \sum_{s \ge 1} \frac{\lambda^s}{s!} T_s,
\end{equation}
where
\begin{equation}
T_s = \sum_{m_1, \cdots, m_s = 1}^{M} \rho_T(\textbf{R}_{m_1}, \cdots, \textbf{R}_{m_2}),
\end{equation}
and where $\rho_T$ gives the truncated $n$-body correlation functions,
\begin{eqnarray}
\rho_T(\textbf{r}_1) &=& \rho(\textbf{r}_1)\nonumber\\
\rho_T(\textbf{r}_1,\textbf{r}_2) &=& \rho(\textbf{r}_1, \textbf{r}_2) - \rho(\textbf{r}_1) \rho(\textbf{r}_2)\\
\rho_T(\textbf{r}_1,\textbf{r}_2,\textbf{r}_3) &=& \rho(\textbf{r}_1,\textbf{r}_2,\textbf{r}_3) - \rho(\textbf{r}_1,\textbf{r}_2) \rho(\textbf{r}_3) - \cdots
\cdots\nonumber
\end{eqnarray}

\subsection{Correlation Functions of the One-Component Plasma}
The exact solution to the one-component plasma found by Jancovici~\cite{jancovici} for coupling parameter $\Gamma = Z^2 e^2/(k_B T) = 2$, where $Z$ is the ion valence and $e$ the elementary charge of an electron.  The $n$-point density correlation functions are given by
\begin{equation}
\rho(\textbf{r}_1, \cdots \textbf{r}_n) = \rho^n \det \left[ e^{- \pi \rho (|z_\mu|^2 + |z_\gamma|^2)/2 + \pi \rho z_\mu \bar{z}_\gamma} \right] |_{\mu, \gamma = 1, \cdots, n}.
\end{equation}
Here, $\rho$ is the background charge density, $z = x + i y$ where $x$ and $y$ describe coordinates on the plane, and $\bar{z}$ its complex conjugate.  After some algebraic manipulation, we can rewrite this expression as
\begin{equation}
\label{eq:correlationfuncs}
\rho(\textbf{r}_1, \cdots \textbf{r}_n) = \rho^n \sum_{\sigma \in S_n} \textrm{sgn}(\sigma) \prod_{j=1}^n \rho_0(\textbf{r}_j - \textbf{r}_{\sigma(j)}) \rho_X (\textbf{r}_j, \textbf{r}_{\sigma(j)}),
\end{equation}
where $S_n$ is the group of permutations on $n$ letters,
\begin{equation}
\rho_0(\textbf{r}) = \exp \left( - \frac{\pi \rho}{2} \textbf{r}^2 \right),
\end{equation}
and
\begin{equation}
\rho_X(\textbf{r}, \textbf{r}') = \exp \left[ - i \pi \rho \left( \textbf{r} \times \textbf{r}' \right) \cdot \hat{z} \right].
\end{equation}
The truncated correlation functions, $\rho_T(\textbf{r}_1, \cdots, \textbf{r}_n)$ are computed by restricting the sum over $\sigma$ in Eq. (\ref{eq:correlationfuncs}) to $n$-cycles.

\begin{figure}[t]
\begin{center}
\resizebox{3in}{!}{\includegraphics{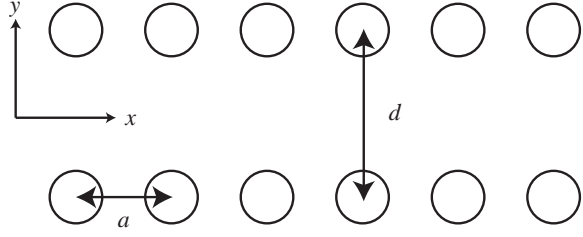}}
\caption{Two infinite lines of adsorption sites separated a distance $d$ and having periodicity $a$.}
\label{fig:picture}
\end{center}
\end{figure}

We suppose that the two lines of sticky sites with periodicity $a$ are separated by a distance $d$ (see Fig.~\ref{fig:picture}).  We describe the sticky site locations by introducing integer variables $n_i$ and Ising variables $\delta_i$ that take on a value of either $0$ or $1$.  Then any $\textbf{R}_i$ can be written as
\begin{equation}
\label{eq:vectors}
\textbf{R}_i = a n_i \hat{x} + d \delta_i \hat{y}.
\end{equation}

Since we are calculating the sums over all the positions of the particles, all $n$-cycles are equivalent by a suitable relabeling of the summation indices.  This leads to the general expression
\begin{eqnarray}
T_s &=& \rho^s (s-1)! (-1)^s \sum_{n_1, \cdots} \sum_{\delta_1, \cdots} \rho_0(\textbf{R}_{m_1,\delta_1}-\textbf{R}_{m_2,\delta_2})  \rho_0(\textbf{R}_{m_2, \delta_2}-\textbf{R}_{m_3},\delta_3) \cdots \\
& & \times \rho_0(\textbf{R}_{m_s,\delta_s} - \textbf{R}_{m_1,\delta_1}) \rho_X(\textbf{R}_{m_1,\delta_1}, \textbf{R}_{m_2,\delta_2}) \rho_X(\textbf{R}_{m_2,\delta_2}, \textbf{R}_{m_3,\delta_3}) \cdots\nonumber\\
& & \times \rho_X(\textbf{R}_{m_s,\delta_s}, \textbf{R}_{m_1,\delta_1}).\nonumber
\end{eqnarray}

\section{Free energy}
We first consider $\rho_X(\textbf{R}_{1},\textbf{R}_{2})$, and substitute in the adsorption site positions from Eq. (\ref{eq:vectors}).  This gives
\begin{equation}
\rho_X(\textbf{R}_1, \textbf{R}_2) = e^{- i \pi \rho a d (n_1 \delta_2 - \delta_1 n_2)}.
\end{equation}
For the particular choice of background charge density
\begin{equation}
\label{eq:densities}
\rho = 2 m/(a d)
\end{equation}
 where $m$ is a positive integer, $\rho_X = 1$ when evaluated on the adsorption sites.  We therefore specialize to densities where this simplification occurs.

We also find that $(\textbf{R}_1 - \textbf{R}_2)^2 = a^2 (n_1 - n_2)^2 + d^2 (\delta_1 - \delta_2)^2$, leading to
\begin{eqnarray}
T_s &=& \rho^s (s-1)! (-1)^s \sum_{n_1, \cdots} e^{- t [(n_1 - n_2)^2 + \cdots + (n_s - n_1)^2]}\nonumber\\
& & \sum_{\delta_1, \cdots} e^{- t' [(\delta_1 - \delta_2)^2 + \cdots + (\delta_s - \delta_1)^2]},
\end{eqnarray}
where $t = \pi \rho a^2/2$ and $t' = \pi \rho d^2/2$.  Since the sum over the $n_i$ and the sum over the $\delta_i$ decompose, we can evaluate the sum over $\delta_i$ using a transfer matrix.  We define the transfer matrix, $\mathcal{T}$, to have the components
\begin{equation}
\mathcal{T} = \left(
\begin{array}{c c}
1 & e^{- \pi \rho d^2/2}\\
e^{- \pi \rho d^2/2} & 1
\end{array}
\right).
\end{equation}
Then
\begin{equation}
T_s = \rho^s (s-1)! (-1)^s  \textrm{tr}(\mathcal{T}^s) \sum_{n_1, \cdots} e^{- t [(n_1 - n_2)^2 + \cdots + (n_s - n_1)^2]}.
\end{equation}
Diagonalizing $\mathcal{T}$ gives the eigenvalues $\lambda_\pm = 1 \pm e^{- \pi \rho d^2/2}$, allowing us to take the trace easily.  Notice that the decoupling of the Ising variables, $\delta_i$ and integer variables, $n_i$ only decouple at densities given by Eq. (\ref{eq:densities}).  At other densities, these additional couplings between the $n_i$ and $\delta_i$ complicates the evaluation of the transfer matrix trace.

The sum over the $n_i$ can be expressed in terms of Jacobi theta functions~\cite{localizedadsorption}, where the Jacobi theta function is defined as
\begin{equation}
\theta_3(\zeta, t) = \sum_{n=-\infty}^{\infty} e^{-t n^2} e^{2 \pi i n \zeta}.
\end{equation}
First, notice that
\begin{equation}
\int d\zeta~\theta_3^s(\zeta,t) = \sum_{n_1, \cdots, n_{s-1}} e^{-t [n_1^2 + n_2^2 + \cdots + (n_1+n_2 + \cdots)^2]}
\end{equation}
for $s>1$ and $\int d\zeta~\theta_3(\zeta,t) = 1$.  This leads to the expression
\begin{equation}
T_s = \rho^s (s-1)! (-1)^s \left[ (1+e^{-\pi \rho d^2/2})^s + (1-e^{- \pi \rho d^2/2})^s \right] \int_0^1 d\zeta~\theta_3(\zeta,\pi \rho a^2/2)^s,
\end{equation}

Substituting this into equation (\ref{eq:freeenergysum}) leads to a sum of two series, both of which are absolutely summable when $| \lambda \rho (1+e^{- \pi \rho d^2/2}) \theta(0,t) | < 1$.  By analytic continuation, we extend this sum to the full range of parameters, leading to the free energy difference between the OCP with and without adsorption sites given by

\begin{eqnarray}\label{eq:free}
\Delta f = - \frac{k_B T}{a^2} \int_0^1 d\zeta~\ln \left[\left\{1+\lambda \rho e^{-( \pi \rho d^2/2)} \theta_3(\zeta, \pi \rho a^2/2)\right\}^2 -\lambda^2 \rho^2 e^{- \pi \rho d^2}\theta_3^2(\zeta, \pi \rho a^2/2) \right] .\nonumber\\
\end{eqnarray}

\begin{figure}[t]
\begin{center}
\resizebox{3in}{!}{\includegraphics{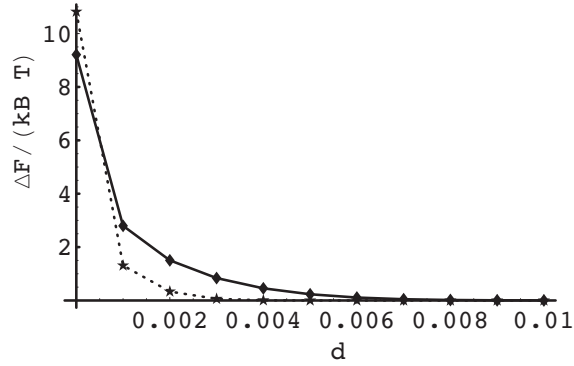}}
\caption{A typical set of free energies $\Delta f - \Delta f_{d \rightarrow \infty}$ with background density $\rho a d_0/2 = 1$ (circles, solid lines) and $\rho a d_0/2 = 5$ (stars, dotted lines).  Here, $d$ is measured in units of $a$ and is an integer multiple of $0.1 a$.  We further set $\lambda a^2 = 100$.  The lines are guides to the eye.}
\label{fig:energy1}
\end{center}
\end{figure}

\begin{figure}[t]
\begin{center}
\resizebox{3in}{!}{\includegraphics{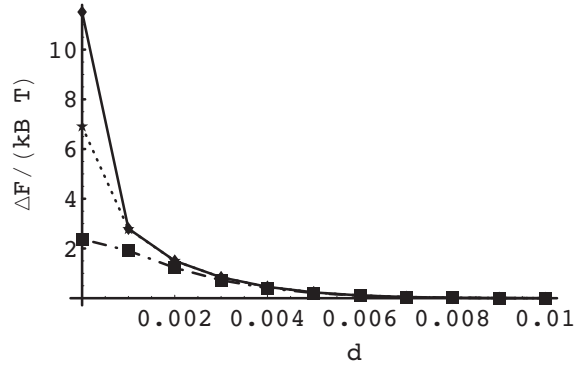}}
\caption{A typical set of free energies $\Delta f - \Delta f_{d \rightarrow \infty}$ with background density $\rho a d_0/2 = 1$ for values of $\lambda a^2 = 0.001$ (squares, dashed line), $0.1$ (stars, dotted line), $10$ (circles, solid line).   Here, $d$ is measured in units of $a$ and is an integer multiple of $0.1 a$.  The lines are guides to the eye.}
\label{fig:energy2}
\end{center}
\end{figure}

In the limit that $d \rightarrow \infty$, we expect the free energy to be a sum of the free energies of two independent lines of sticky sites.  Indeed, this limit yields
\begin{equation}
\Delta f_{d \rightarrow \infty} = - 2 \frac{k_B T}{a^2} \int_0^1 d\zeta~\ln \left[1+\lambda \rho \theta_3(\zeta, \pi \rho a^2/2) \right].
\end{equation}
In the opposite limit, $d \rightarrow 0$, we expect the free energy to agree with that of a single line of adsorption sites with a potential given by $2 \lambda$.  It is easy to see that the free energy in this limit is
\begin{equation}
\Delta f_{d \rightarrow 0} =  - \frac{k_B T}{a^2} \int_0^1 d\zeta~\ln \left[1+2 \lambda \rho \theta_3(\zeta, \pi \rho a^2/2) \right].
\end{equation}

Written in terms of the dimensionless constants $t = \pi \rho a^2/2 = \pi a/d$, $t' = \pi \rho d^2/2 = \pi d/a$ and $\Lambda = \lambda \rho = 2 \lambda/(a d)$, we find the change in free energy as the adsorption sites approach each other, $\Delta F = \Delta f - \Delta f_{d \rightarrow \infty}$, to be given by
\begin{eqnarray}
\Delta F &=& - \frac{k_B T}{a^2} \int_0^1 d\zeta~\ln \left[ \frac{1+\Lambda (1+e^{- t'}) \theta_3(\zeta, t)}{1+\Lambda \theta_3(\zeta,t)} \right]\\
& &- \frac{k_B T}{a^2} \int_0^1 d\zeta~\ln \left[ \frac{1+\Lambda (1-e^{- t'}) \theta_3(\zeta,t)}{1+\Lambda \theta_3(\zeta,t)} \right].\nonumber
\end{eqnarray}

Recall that this free energy is only valid when $\rho a d/2= m$ for any positive integer $m$.  Thus, we can compare the free energy of states with the same lattice constant $a$ and background charge density $\rho$, only for integer multiples of some specific valid separation $d$.  To be more specific, suppose we have the free energy at some density such that $\rho a d/2 = 1$ and separation $d$, then at separation $m d$ in the same background density, we have $\rho a d/2 = m$.  Thus, the free energy formula Eq. (\ref{eq:free}) will be valid only for integer multiples of $d$.

\begin{figure}[t]
\begin{center}
\resizebox{3in}{!}{\includegraphics{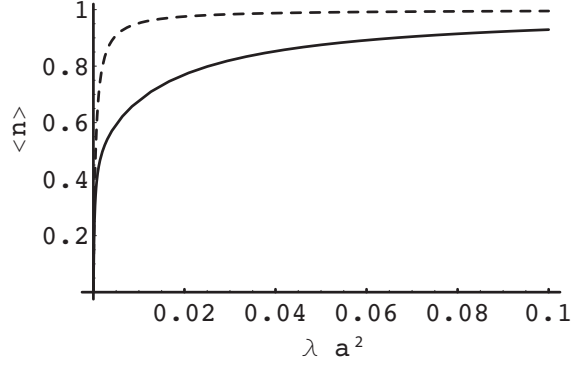}}
\caption{The fraction of occupied sites as a function of $\lambda a^2$ for two lines separated by a distance $d = 0.1 a$ with $\rho a d/2 = 1$ (solid).  This is compared to the fraction of occupied sites for a single line of adsorption sites at the same background density (dashed line).}
\label{fig:avgn}
\end{center}
\end{figure}

In Fig.~\ref{fig:energy1}, we plot $\Delta F$ as a function of $d$ for two values of the background density, $\rho$, given by $\rho a d_0/2 = 1$ and $\rho a d_0/2 = 5$, where $d_0 = 0.1 a$ is the smallest value of $d$.  The lines of adsorption sites are always repulsive.  For larger $d$, $\Delta F$ becomes zero quickly.  In Fig.~\ref{fig:energy2}, we plot $\Delta F$ as a function of $d$ for different values of $\lambda a^2$.  As the depth of the potential is increased, the repulsive strength of the interaction is also increased.  This is indicative of the adsorption sites pinning charge on them, leading to their repulsion.

Finally, we compute the average occupation number of a site.  This is given by~\cite{localizedadsorption}
\begin{equation}
\langle n \rangle = - a^2 \lambda \frac{\partial}{\partial \lambda} \frac{\Delta f}{k_B T}.
\end{equation}
This is plotted as a function of $\lambda a^2$ for in Fig.~\ref{fig:avgn} for two lines with separation $d = 0.1 a$ and $\rho a d/2 = 1$ (solid line) and for a single line of adsorption sites (dashed line).

It is clear from Fig.~\ref{fig:avgn} that the repulsion inhibits the adsorption of the ions.  However, the separation $d=0.1 a$ is very small.  At separations on the order of the site spacing, there is no appreciable difference in the fraction of occupied sites as a function of $\lambda$.

\section{Average density}
In this section, we will compute the average counterion density at an arbitrary point $\textbf{R}_0$.  We can find the density directly by fixing the position of one of the ions.  This will require the computation of the quantity
\begin{equation}
\label{eq:Tdensity}
T_s(\textbf{R}_0) = \rho^{s+1} \sum_{\textbf{R}_1} \cdots \sum_{\textbf{R}_s} \prod_{j=0}^s \rho_0(\textbf{R}_j,-\textbf{R}_{\sigma(j)}) \rho_X(\textbf{R}_j, \textbf{R}_{\sigma(j)}),
\end{equation}
where the sum is over all $s$-cycles $\sigma$.  Then the average density is given by~\cite{localizedadsorption}
\begin{equation}
\label{eq:density}
\langle \rho(x_0,y_0) \rangle - \rho = \sum_{s=1}^{\infty} \frac{\lambda^s}{s!} T_s(\textbf{R}_0)
\end{equation}
where $x_0$ and $y_0$ are the components of $\textbf{R}_0$.

For $\pi \rho a d = 2 \pi n$, $n$ an integer, and for $\textbf{R}_j = a n_j \hat{x} + d \delta_j \hat{y}$, equation (\ref{eq:Tdensity}) decomposes into the product
\begin{equation}
T_s = \rho^{s+1} s! (-1)^s T_s^{(1)} T_s^{(2)},
\end{equation}
where
\begin{eqnarray}
T_s^{(1)} &=& \sum_{n_1 \cdots n_s} e^{- \pi \rho x_0^2/2} e^{- (\pi \rho a^2/2) [ n_1^2 + (n_1-n_2)^2 + \cdots + (n_{s-1} - n_s)^2 + n_s^2]}\nonumber\\
& & \times e^{- \pi \rho a x_0 (n_1 + n_s) + i \pi \rho a y_0 (n_1 - n_s)}
\end{eqnarray}
and
\begin{eqnarray}
T_s^{(2)} &=& \sum_{\delta_1 \cdots \delta_s} e^{- \pi \rho y_0^2/2} e^{- (\pi \rho d^2/2) [\delta_1^2 + (\delta_1 - \delta_2)^2 + \cdots +\delta_s^2]}\nonumber\\
& & \times e^{- \pi \rho d y_0 (\delta_1 + \delta_s)} e^{i \pi \rho d (\delta_1 - \delta_s)}.
\end{eqnarray}

Defining $z = x_0 + i y_0$ and $\bar{z}$ its complex conjugate, we find that
\begin{equation}
T_s^{(1)} = \sum_{n_1 \cdots n_s} e^{- \pi \rho x_0^2/2} e^{- t [n_1^2 + \cdots + n_s^2 + (n_1 + \cdots n_s)^2]} e^{- \pi \rho a (n_1 z - n_s \bar{z})}.
\end{equation}
Using the transfer matrix $\mathcal{T}$ and defining a new diagonal matrix,
\begin{equation}
\mathcal{M} =  \left(
\begin{array}{c c}
1 & 0\\
0 & e^{- \pi \rho d^2/2 + \pi \rho d y_0 + i \pi \rho d x_0}
\end{array}
\right),
\end{equation}
and its complex conjugate $\bar{\mathcal{M}}$, we find
\begin{equation}
T_s^{(2)} = e^{- \pi \rho y_0^2/2} \textrm{tr}(\mathcal{T}^{s-1} \bar{\mathcal{M}} T \mathcal{M})
\end{equation}
for $s \ge 1$ and $T_0^{(2)} = e^{- \pi \rho y_0^2/2}$.
Evaluating $T_s^{(2)}$ gives
\begin{equation}
T_s^{(2)} = \left[ \lambda_+^{s-1} G_+(z) + \lambda_-^{s-1} G_-(z) \right],
\end{equation}
where $4 G_+(z) = \exp(- \pi \rho y_0^2/2) (\lambda_+ |1+A|^2 + \lambda_- |1-A|^2)$ and $4 G_-(z) = \exp(- \pi \rho y_0^2/2) (\lambda_- |1+A|^2 + \lambda_+ |1-A|^2)$, and $A = \exp (- \pi \rho d^2/2 + \pi \rho d y_0 + i \pi \rho d x_0)$.

\begin{figure}[t]
\begin{center}
\resizebox{3in}{!}{\includegraphics{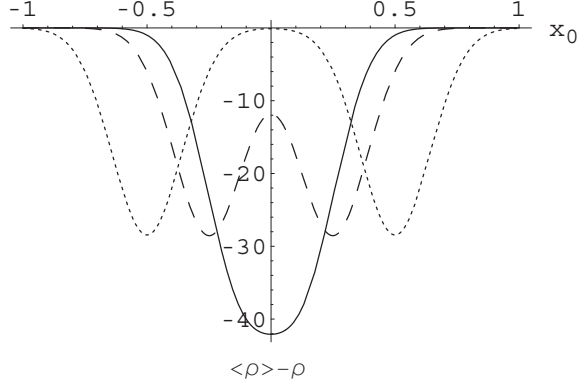}}
\caption{$\langle \rho \rangle - \rho$ as a function of $y_0$ for $x_0=0$ in units of $a$.  Here $\rho a d = 2 n$ and $\lambda \rho = 8$.  We show $d=0.25 a$ (solid), $d=0.5 a$ (dashed), and $d=1 a$ (dotted).}
\label{fig:density}
\end{center}
\end{figure}

Using the results of Rosinberg \textit{et al.}~\cite{localizedadsorption}, we can compute $T_s^{(2)}$ in terms of $F(\zeta,z,t) = \{ \theta_3[\zeta+i \bar{z} t/(\pi a),t] \theta_3[\zeta+i z t/(\pi a),t] + \theta_3[\zeta- i \bar{z} t/(\pi a),t] \theta_3[\zeta-i z t/(\pi a),t] \}/2$.  This gives
\begin{equation}
T_s = (-1)^s s! \rho^{s+1} \left[ \lambda_+^{s-1} G_+(z) + \lambda_-^{s-1} G_-(z) \right] e^{- \pi x_0^2/2} \int_0^1 d\zeta~\theta_3^{s-1}(\zeta,t) F(\zeta,z,t).
\end{equation}

Using equation (\ref{eq:density}), we find the average density
\begin{eqnarray}
\langle \rho(z) \rangle - \rho &=& - \lambda G_+(z) \rho^2 e^{- \pi \rho x_0^2} \int_0^1 d\zeta~\frac{F(\zeta,z,t)}{1+\lambda \rho \lambda_+ \theta_3(\zeta,t)}\nonumber\\
& & - \lambda G_-(z) \rho^2 e^{- \pi \rho x_0^2} \int_0^1 d\zeta~\frac{F(\zeta,z,t)}{1+\lambda \rho \lambda_- \theta_3(\zeta,t)}.
\end{eqnarray}
Finally, we note that
\begin{eqnarray}
4 G_+(x_0,y_0) &=& \lambda_+ \left\{ e^{- \pi \rho y_0^2} + e^{- \pi \rho (y_0-d)^2} + 2 e^{- \pi \rho [y_0^2 + (y_0-d)^2]/2} \cos ( \pi \rho d x_0 ) \right\}\\
& & + \lambda_-  \left\{ e^{- \pi \rho y_0^2} + e^{- \pi \rho (y_0-d)^2} - 2 e^{- \pi \rho [y_0^2 + (y_0-d)^2]/2} \cos ( \pi \rho d x_0 ) \right\},\nonumber
\end{eqnarray}
and
\begin{eqnarray}
4 G_-(x_0,y_0) &=& \lambda_- \left\{ e^{- \pi \rho y_0^2} + e^{- \pi \rho (y_0-d)^2} + 2 e^{- \pi \rho [y_0^2 + (y_0-d)^2]/2} \cos ( \pi \rho d x_0 ) \right\}\\
& & + \lambda_+  \left\{ e^{- \pi \rho y_0^2} + e^{- \pi \rho (y_0-d)^2} - 2 e^{- \pi \rho [y_0^2 + (y_0-d)^2]/2} \cos ( \pi \rho d x_0 ) \right\}.\nonumber
\end{eqnarray}

Notice that, for $\pi \rho a d = 2 \pi n$, the periodicity of $G_\pm(x_0,y_0)$ in $x_0$ is always commensurate with the lattice spacing $a$.  In Fig.~\ref{fig:density}, we plot $\langle \rho \rangle - \rho$ as a function of $y_0$ for $x_0 = 0$, $\rho a d = 2 n$, and $\lambda \rho = 8$ and for a variety of different spacings.  Notice that the density has two peaks for large separations but a single peak as the separation $d$ becomes smaller than the periodicity $a$.  As a function of $x_0$, the density is always periodic with period $a$.

\section{Discussion}
The main conclusion of our calculation is that for the geometry that we have chosen no attractive forces are induced by the discrete structure of the charged line. \\

The derivation of $\Delta F$ is  valid for $d$ taking values that are integer multiples of $2/(a \rho)$.  Further, Eq. (\ref{eq:free}) gives the correct free energy for $d = 0$ (as we have already seen).  It is conceivable, then, that Eq. (\ref{eq:free}) is correct for all values of $\rho$ and $d$.

One of the features of this model that makes an exact evaluation possible is that a density $\rho$ can be found such that the Ising variables $\delta_i$ and the integer variables $n_i$ are uncoupled.  When $\rho a d \ne 2 m$ for an integer $m$, a coupling does indeed arise that makes the computation of the free energy more difficult.  Additionally, if the adsorption sites are not aligned, an additional $\hat{x}~\delta_i~\Delta a$ component arises in $\textbf{R}_i$, where $\Delta a$ measures the degree of misalignment.  This component introduces a coupling between the $n_i$ and $\delta_i$ that will be discussed in future work.
\section{Acknowledements}
L.B. wants to acknowledge the warm hospitality of Fyl Pincus at the MRL of the University of California in Santa Barbara, where this project was started. The idea of this research was inspired during discussions with Fyl.  Support
from NSF through grant DMR02-03755 and DOE  grant DE-FG02-03ER 15422 is also graciously acknowledged.\\

\end{document}